# Angular-spectrum-dependent interference

Chen Yang[1,2], Zhi-Yuan Zhou[1,2]*, Yan Li[1,2], Shi-Kai Liu[1,2], Zheng Ge[1,2], Guang-Can Guo[1,2] and Bao-Sen Shi[1,2]*

[1] CAS Key Laboratory of Quantum Information, University of Science and Technology of China, Hefei, Anhui 230026, China
[2] Synergetic Innovation Center of Quantum Information & Quantum Physics, University of Science and Technology of China, Hefei, Anhui 230026, China
* Correspondence: (Zhi-Yuan Zhou) zyzhouphy@ustc.edu.cn or (Bao-sen Shi) drshi@ustc.edu.cn

**Abstract**

Optical interference is not only a fundamental phenomenon that has enabled new theories of light to be derived but it has also been used in interferometry for the measurement of small displacements, refractive index changes and surface irregularities. In a two-beam interferometer, variations in the interference fringes are used as a diagnostic for anything that causes the optical path difference (OPD) to change; therefore, for a specified OPD, greater variation in the fringes indicates better measurement sensitivity. Here, we introduce and experimentally validate an interesting optical interference phenomenon that uses photons with a structured frequency-angular spectrum, which are generated from a spontaneous parametric down-conversion process in a nonlinear crystal. This interference phenomenon is manifested as interference fringes that vary much more rapidly with increasing OPD than the corresponding fringes for equal-inclination interference; the phenomenon is parameterised using an equivalent wavelength, which under our experimental conditions is 29.38 nm or about 1/27 of the real wavelength. This phenomenon not only enriches the knowledge with regard to optical interference but also offers promise for applications in interferometry.

**Introduction**

Since the observation of double-slit interference by Young in 1807, optical interference phenomena have provided multiple demonstrations of the wave nature of light. After that pioneering experiment, many studies on interferences have been performed to reveal the deeper nature of light, for example, the wave–particle duality of photons[1-3] and their high-order correlations[4,5]. To date, interference phenomena have been observed not only in the light intensity, but also in other degrees of freedom of light[6], including the frequency[7], polarisation[8] and orbital angular momentum[9], and have thus played an important role in various structured light generation applications[10]. Fringe patterns are a common feature of most interference phenomena and these fringes form the basis of interferometers, which have proven to be powerful practical tools in numerous fields, e.g., in gravitational-wave detection[11], optical coherence tomography[12], Fourier transform infrared spectroscopy[13], and applications of fibre optic gyroscopes[14].

For light intensity interference, the existence of constructive and destructive interference is dependent on a stable phase difference between two or more light beams. In traditional interferometers, the stable phase difference is determined by the optical path difference (OPD). For example, in equal-inclination interference[15] (Fig. 1a), the OPD between the two reflecting surfaces changes with incident angle, and therefore, light with the same incident angle finally superposes to form a bright or dark fringe. Features of the interference fringe patterns are also dependent on the properties of the light source. Most past studies and applications of interference have used lasers or thermal light sources. In recent years, a new light source based on spontaneous parametric down-conversion (SPDC)[16,17] in nonlinear crystals has been attracting much attention. The SPDC is a second-order nonlinear process, in which a higher-energy pump photon splits into a pair of lower-energy photons, one designated a signal photon and the other an idler photon, emerging with a certain probability from a nonlinear crystal. This special source of light has helped in finding many novel interference phenomena[4,18-20] and applications[21-30] that are quite different from those using lasers or thermal light sources[31-33]. If the entanglement properties are ignored, each arm (subsystem) of an SPDC source can usually be regarded as an incoherent mix of photons with all possible spatial modes and frequencies. Unlike lasers or thermal light sources, in which the spatial modes and frequency components can be treated independently, photons from an SPDC source have a structured frequency-angular spectrum (FAS) caused by the phase-matching conditions. The emission angles outside the nonlinear crystal are dependent on the emitted photon frequencies. For a long crystal, this dependence relation is approximately a one-to-one mapping that is governed by a tuning curve[31], which can be approximated as a parabola.

In this work, we have observed a distinctive two-beam interference phenomenon in an amplitude division interferometer using photons from one arm of an SPDC source (Fig. 1b); we refer to it as angular-spectrum-dependent (ASD) interference because it is caused by a combination of interference patterns of different angular components. The principle and phenomenon of the ASD interference are very similar to those of the traditional equal-inclination interference: they both have ring-like fringes, the phase difference inducing bright or dark rings is dependent on the angle, and the number of rings is dependent on the distance $d$. However, ASD interference is fundamentally different from traditional equal-inclination interference. The creation and properties of ASD interference are closely related to the frequency-angular one-to-one mapping relation of the SPDC process. To illustrate the properties of ASD interference (Fig. 1b) and distinguish it from the traditional equal-inclination interference (Fig. 1a), we compare them in terms of the following five aspects. First, the two light sources have different radiation properties: the point source shown in Fig. 1a radiates spherical waves that are isotropic, but the SPDC process shown in Fig. 1b radiates photons over a very wide spectrum, where the photon frequencies are related to the emission angle $\theta(\omega)$, which is shown using Eq. (1). Second, with regard to their principles of interference, the phase differences $n\pi$ for the bright or dark fringes are caused by the angular-dependent OPDs $\Delta(\theta)$ of the light in Fig. 1a, whereas the phase differences $n\pi$ are caused by the specific photon frequencies $\omega_n$ in Fig. 1b. Third, in Fig. 1a, the photons in each of the fringes are coherent and have the same spectrum and the interference visibility is thus dependent on the width of the spectrum; in Fig. 1b, however, the photons in the different fringes have different frequencies and the fringe visibility is dependent on the width of the FAS. Fourth, in the optical setups, the lens in Fig. 1a allows observation of the far-field of the fringes that are created, while the lens shown in Fig. 1b is used for collimation. Finally, the interference patterns of the two phenomena are both ring-like fringes, but with increasing distance $d$ between the two reflecting surfaces, the fringes of the ASD interference vary much more quickly than those in the traditional interference pattern; in other words, much shorter distance $d$ are required for the ASD interference to obtain the same interference patterns. In stressing this last point, we say that the equivalent wavelength of this ASD interference is much shorter than the actual wavelength. The physical meaning of the equivalent wavelength here is that the ASD interference fringes are the same as those from a traditional equal-inclination interferometer in which the wavelength of the photons has this value. In the following, the equivalent wavelength is defined so that the expression for the phase difference has the same form as that for the traditional interference.

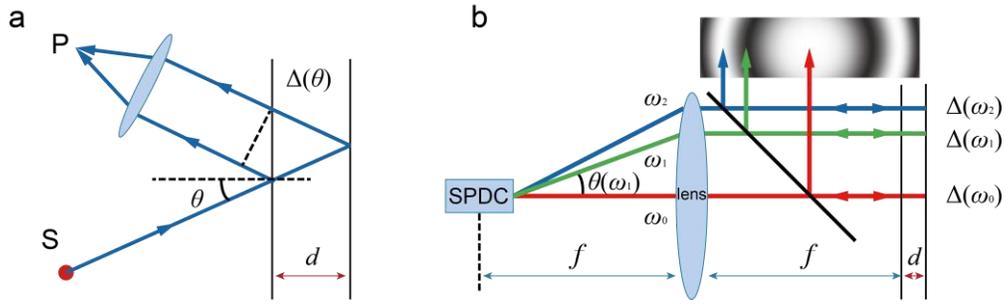

**Fig. 1 | Schematics**: **a** Traditional equal-inclination interference. Two rays from a point source S interfere at the point P. The phase differences for the bright or dark fringes are caused by the angular-dependent OPDs $\Delta(\theta)=2d\cos\theta$ (where $\Delta$ denotes an OPD). **b** Angular-spectrum-dependent interference. The phase differences for each fringe order are caused by the frequencies $\omega_n$ (where $\omega$ denotes a frequency). The OPD is given by $\Delta(\omega_n)=2d\omega_n/c$.

The FAS of the SPDC has been reported previously[31,34]. Shih calculated the tuning curve required for type-I and type-II angle phase-matching[31]. Burlakov et al.[34] calculated the intensity distribution of the FAS near the degenerate phase-matching condition and presented a photograph of this distribution; they implemented the second-order and fourth-order interference using photons from two nonlinear interaction regions, however, only single-frequency interference was observed in their experiment. Nevertheless, the ring-like fringes created by interference using photons with the structured FAS remain unexplored, along with the properties of these fringes, and these fringes thus form the main topic of our study. We also quantify the distribution of the fringes and their differences from the fringes obtained through traditional equal-inclination interference. In the following, we first introduce briefly the experimental setup (details are presented in Methods), describe the FAS of the SPDC obtained from our experiment and present the expression for the tuning curve used for nondegenerate type-0 quasi-phase-matching. Next, we explain how the interference fringes are generated and define the equivalent wavelength and parameter $\gamma$, which is the ratio of the real centre wavelength to the equivalent wavelength, to show the difference between ASD and equal-inclination interference. Finally, we discuss the potential applications of this ASD interference phenomenon.

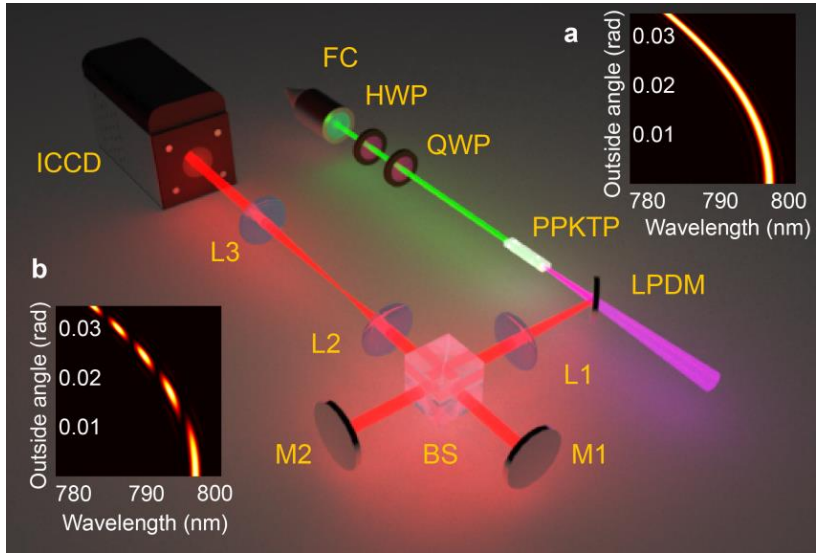

**Fig. 2 | Schematic of the interferometer**. The focal lengths of lens L1–L3 are 100 mm, 100 mm, and 200 mm. M1, M2: mirrors; BS: beam splitter; LPDM: long-pass dichroism mirror; FC: fibre collimator; HWP and QWP are half waveplate and quarter waveplate; PPKTP: periodically poled potassium titanyl phosphate; ICCD: intensified charge-coupled device. Insets: frequency-angular spectrum of (**a**) the signal photons outside the crystal and (**b**) after the interferometer and lens L2. The frequency-angular spectrums in (a) and (b) are obtained by simulation using equations (S4) and (S5) in the Supplementary Information, respectively.

## Results

In our experiment (Fig. 2), we use a periodically poled potassium titanyl phosphate (PPKTP) crystal as our SPDC source, which makes use of nondegenerate type-0 quasi-phase-matching[35]. The photons generated in SPDC and used for the interference are referred to as signal photons (with wavelengths of approximately 797 nm); the idler photons (~1540 nm) are discarded. The FAS of the signal photon (Fig. 2a) is described by a binary function that reflects the radiation properties of the SPDC process. Its shape is parabolic with a width having a $sinc^2$ functional shape. The unique distribution of FAS is simulated based on the phase-matching condition (see Section 1 of Supplementary Information for details). The function values of the FAS reflect the relative probability of photon detection for a particular outside angle (emission angle outside the crystal) and a particular frequency. If the phase-matching condition is well satisfied, the value is relatively large. In other words, the smaller the phase mismatch is, the larger the value is, and vice versa.

We assume photons with a FAS of Fig. 2a enter a Michelson interferometer having an arm difference $d$. Because each frequency component has a distinct interference result expressed by factor $\left[1+\cos(2d\omega_s/c)\right]/2$, the FAS after the interferometer becomes that shown in Fig. 2b (here, $d=100$ μm as an example). If the photons then pass through a lens that is used as a Fourier translator, each of their angular components maps into a ring in the spatial domain. Therefore, a ring-like interference pattern is formed when the photons are observed. The setup of the Michelson interferometer in our experiment (Fig. 2) comprises two lenses (L1 and L2) that form a 4-f imaging system and two mirrors (M1 and M2) located at the focal points. Another lens (L3) is used as a Fourier translator, which maps the spatial frequency components to the spatial rings on the detection plane. The interference patterns, shown in Fig. 3a, are recorded by a photon-counting intensified charge-coupled device (ICCD) camera. The simulations of the interference patterns from calculating the phase mismatch are shown in Fig.

3b (the simulation is based on equations (S3) and (S6) in Section 2 of the Supplementary Information, in which a small-angle approximation is used).

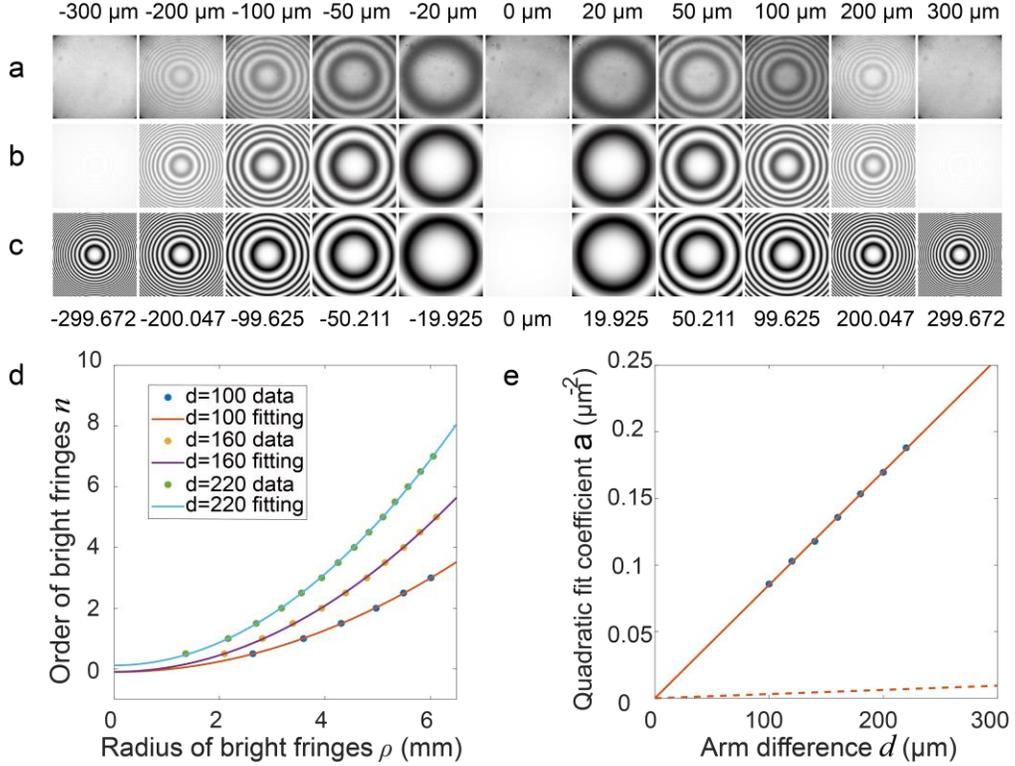

**Fig. 3 | a** Experimental results. **b** Simulation results. **c** Predictions using Eq. (3). **d** The evaluated radial distance on camera of the minima and maxima for $d = 100\,\mu m, 160\,\mu m, 220\,\mu m$. The integer and half-integer orders correspond to maxima and minima, respectively. **e** The quadratic coefficients $a$ are obtained by fitting the data of $d = 100, 120, 140, 160, 180, 200, 220\,\mu m$. The data (blue spots) are fitted by a linear function (red line). The $a_e$-$d$ relation (dashed line) of the equal-inclination interference is shown for comparison.

The interference pattern may also be established from analytical methods. Here, we ignore the width of the FAS (Fig. 2a), and we are only interested in parabola-like tuning curves that show how the outside angle of the signal photons changes as a function of frequency or wavelength. The tuning curve is approximately described by the expressions (detailed derivations may be found in Section 3 of Supplementary)

$$\theta_{s\_out}^2 = b_1 \Delta\omega \tag{1}$$

$$b_1 = \frac{2n_{i0}n_{s0}\omega_{i0}\left(\beta_s\omega_{s0} + n_{s0} - \beta_i\omega_{i0} - n_{i0}\right)}{\omega_{s0}\left(\omega_{s0}n_{s0} + \omega_{i0}n_{i0}\right)} \tag{2}$$

where $\Delta\omega = \omega_s - \omega_{s0} = \omega_{i0} - \omega_i$, and $\beta_s = \left(\frac{dn}{d\omega_s}\right)_{\omega_{s0}}, \beta_i = \left(\frac{dn}{d\omega_i}\right)_{\omega_{i0}}$ are the coefficients of first-order dispersion at the centre frequency of the signal and idler photons. The subscript 0 indicates a corresponding value at the centre frequency of the wavelength; for example, $\omega_{s0}, \omega_{i0}, n_{s0}, n_{i0}$ are the centre frequency and the corresponding refractive index of the signal and idler photons. Equations (1) and (2) were obtained by applying approximate conditions in which $\Delta\omega$ is small and the length of the crystal is long enough so that the width of the FAS may be ignored.

The relationship between radius $\rho$ of the abovementioned ring and the outside angle of signal photons is approximately given by $\rho \approx \theta_{s\_out} f_3$, for which $f_3$ is the focal length of L3. By substituting Eq. (1) into the general interference factor $[1+\cos(2d\omega_s/c)]/2$, the count recorded by the ICCD may be expressed as:

$$C(\rho) \propto 1 + \cos\left(\frac{2d\omega_{s0}}{c} + \frac{2d}{f_3^2 b_1 c}\rho^2\right) \qquad (3)$$

The interference patterns predicted by Eq. (3) are presented in Fig. 3c. The difference between Fig. 3b and c is that the simulations in Fig. 3c do not take the interference visibility into account because the width of the $\text{sinc}^2$ function is ignored. In other words, the complete interference expression should have the form $[1+V(d)\cos(2d\omega/c)]/2$, where the function $V(d)$ represents the interference visibility and $V(d)$ is assumed to have a constant value of 1 in Eq. (3). The interference visibility and the coherence length are described by Eq. (S22) and (S23) in Section 4 of the Supplementary Information, respectively.

In the experiment, we fixed M2 on the displacement platform and M1 on the piezoelectric transducer (PZT). We varied the arm difference by moving M2, then finely adjusted the PZT to ensure the centres of the interference patterns are bright spots. Fig. 3a shows the experimental results with different arm differences. The numbers at the top indicate approximate arm differences, specifically, from the reading of the displacement platform. The numbers at the bottom of Fig. 3c show the actual arm differences set in the simulation. In comparison, the experimental results agree well with our theoretical calculation.

In Fig. 3a–c, more interference rings appear with increasing arm difference $d$. We next show the radial distribution of the rings as a function of the arm difference. Assuming the radius of the $n$-th ring is $\rho_n$, ($n$ being the constructive interference order), then from Eq. (3) we obtain a quadratic relation

$$a\rho_n^2 + \phi_0 = n \qquad (4)$$

where $a = d/\pi f_3^2 b_1 c$ and $\phi_0 = d\omega_{s0}/\pi c$. Eq. (4) describes the distribution of fringes where the coefficient $a$ determines the radius of the fringes for each order. For a specific order $n$, a greater value of $a$ indicates a smaller value of the radius $\rho_n$ and thus indicates a higher fringe density. In the experiments, $a$ is obtained by fitting the experimental data $(\rho_n, n)$. By comparing the coefficient $a_e = d/f^2\lambda$ for the far-field equal-inclination interference[36], ($a_e = d/f_3^2\lambda_{s0}$ for our experimental condition), we define an equivalent wavelength $\lambda_{eq} = \pi b_1 c$ to cause the coefficient $a$ to have the general form $a = d/f_3^2\lambda_{eq}$. In the experiment, this equivalent wavelength is obtained by fitting the dependence of $a$ to $d$. Fig. 3d shows the experimentally obtained pairs $(\rho_n, n)$ for different values of $d$. The coefficient $a$ may be evaluated using a second-order polynomial fit to the data. In Fig. 3e, the obtained values of $a$ are plotted for different $d$. The red line shows the fitted result, from which one obtains the equivalent wavelength 29.38 nm, which agrees well with the predicted value of 29.86 nm. For comparison, we also show the $a_e$-$d$ relation (dashed line) of a traditional equal-inclination interference; the ratio of the slopes of the two lines is denoted $\gamma = a/a_e = \lambda_{s0}/\lambda_{eq}$. Except for the centre wavelength, $\gamma$ is also dependent on the key parameter $b_1$, the value of which depends on properties of the crystal material. From a qualitative analysis using Eq. (2), the determining factor for $b_1$ includes the refractive index and crystal dispersion, the degree of degeneracy, and the type of quasi-phase-matching. The value of $\gamma$ can be larger if the experimental parameters are carefully selected.

**Discussion**

In summary, we report and study an interference phenomenon known as ASD interference using photons from one arm of an SPDC source. In this type of interference, the fringes distribution in Eq. (4) is the same as that in equal-inclination interference, however, it varies more rapidly with the increasing interferometer arm difference than those obtained from traditional equal-inclination interference. We defined two parameters to quantitatively compare the difference between the ASD interference and the traditional equal-inclination interference: the equivalent wavelength $\lambda_{eq}$ and the ratio $\gamma$. Under our experimental conditions, $\gamma$ has an approximate value of 27; this means that the fringe density is improved by 27-fold for a specific arm difference $d$, in other words, the fringes of this interference vary 27 times more rapidly than the traditional equal-inclination interference with increasing arm difference. An advantage of the ASD interference is that the sensitivity can be increased $\gamma$-fold when we use this interferometer to measure small displacements or refractive index changes by recording variations of fringes, because in these cases, greater variation in the fringes indicates better sensitivity for OPD.

Another advantage of ASD interference with large value of $\gamma$ is that the point at which the zero OPD occurs can be determined more accurately and thus the optical path measurement accuracy can be improved. As shown in Fig. 3a–c, the first completely dark fringe occurs when $d = \pm 20\,\mu m$; this means that the position with the equivalent path can be determined with an error of $\pm 20\,\mu m$; in the supplementary, we show that the error can be reduced to $\pm 0.54\,\mu m$ by fitting our experimental data. Furthermore, Eq. (3) indicates that the accuracy may be improved further by expanding the field of view $\rho_{max}$ or reducing either the focal length of L3 or the equivalent wavelength $\lambda_{eq}$. Because the SPDC source itself is a currently available nondegenerate two-photon source, the potential applications of ASD interference can also be generalised to the nonlinear interferometers based on SPDC[34,37,38].

The ASD interference fringes not only have a ring-like structure in intensity but also have a structure in frequency of photons, where the photons in the different fringes have different frequencies. Considering the frequency structure, this interference phenomenon also holds promise in spectral-shaping a photon source based on SPDC. Because the rings in the interference patterns map different wavelength components, a cosine-modulated frequency spectrum is obtained if the interference patterns are collected into multimode fibres.

The interference phenomenon can be used in a reverse manner to measure the tuning curve of the SPDC process. By fitting the equivalent wavelength, the parameter $b_1$ of the tuning curve can then be obtained. Overall, the novel phenomenon reported here not only enriches the existing knowledge with regard to interference and SPDC but also has promise for use in interferometry applications.

**Materials and methods**
**Pump laser.**
The 525.2-nm light beam of the CW pump laser is generated in single-pass sum-frequency generation (SFG) with a 10-mm type-0 periodically poled potassium titanyl phosphate (PPKTP) crystal (the SFG source is omitted in Fig. 2). In the SFG source, the wavelengths of the two pump beams are 1540 nm and 797 nm, and all three beams are vertically polarised. The SFG laser beam is collected into a single-mode fibre and exits through a fibre collimator (the FC in Fig. 2). The idealised plane-wave pump in the SPDC leads to strict transverse momentum correlations. Therefore, the pump beam is collimated by a lens group (the lens group is omitted in Fig. 2); its width is of order 400 μm and the pump power is 50 mW. The waveplates (Fig. 2) are used to transform the pump beam from the collimator into a vertically polarised beam.

### Crystals.
Two PPKTP crystals are used in the experiment, one for SFG and the other for SPDC. The two crystals have the same parameter values: their dimensions are 1 mm × 2 mm × 10 mm, and their grating periods are 9.34 μm. The temperature of the crystal used during SFG is set at 24 ℃, which is an optimum temperature for SFG. The temperature of the crystal used for SPDC is set at 29 ℃. This temperature is determined by performing difference-frequency generation between the 525.2-nm and 1540-nm laser beams. The two temperatures are different because the widths of the beams in the two crystals are different.

### Optical setup.
Because the SPDC is the inverse of SFG, the central wavelength of the idler and signal photons are approximately 1540 nm and 797 nm. The signal and idler photons are split through a long-pass dichroic mirror (DM), where the idler photons pass through the DM (discarded) and the signal photons are reflected. The pump beam is filtered by a 750-nm long-pass filter that is omitted in Fig. 2. The experiment was performed in a dark environment, and the light path in front of the camera was carefully shaded by a sealed box to block external light.

### Data acquisition.
The interference patterns (Fig. 3a) were recorded by our ICCD camera (Andor iStar DH334T) with a 10 s exposure time. The working temperature of the ICCD is cooled at −25 °C. The background is taken before the data acquisition and is subtracted by the ICCD camera automatically when signals are taken. The average counts of each pixel of the background are around 9000. The normalised grey values of the images in Fig. 3a from 0 to 1 represent the counts from 0 to 10000.

### Acknowledgements
This work was supported by the National Natural Science Foundation of China (NSFC) (61605194, 11934013, 61525504), the Anhui Initiative In Quantum Information Technologies (AHY020200), the China Postdoctoral Science Foundation (2017M622003, 2018M642517).

### Conflict of interests
The authors declare that they have no conflict of interest.

### Contributions
C.Y., Z.-Y.Z. and B.-S.S. conceived the research and designed the experiments. C.Y. performed the experiments and numerical simulations. The data acquisition and processing were performed by C.Y. with help from Y.L., S.-K.L., and Z.G., Z.-Y.Z., G.-C.G, and B.-S.S. supervised the project. All authors contributed to the discussion of experimental results. C.Y., Z.-Y.Z. and B.-S.S. wrote the manuscript with contributions from all co-authors.

# Supplementary Information for

# Angular-spectrum-dependent interference

Chen Yang[1,2], Zhi-Yuan Zhou[1,2]*, Yan Li[1,2], Shi-Kai Liu[1,2], Zheng Ge[1,2], Guang-Can Guo[1,2] and Bao-Sen Shi [1,2]*

[1] CAS Key Laboratory of Quantum Information, University of Science and Technology of China, Hefei, Anhui 230026, China

[2] Synergetic Innovation Center of Quantum Information & Quantum Physics, University of Science and Technology of China, Hefei, Anhui 230026, China

* Correspondence: (Zhi-Yuan Zhou) *zyzhouphy@ustc.edu.c*n or (Bao-Sen Shi) *drshi@ustc.edu.cn*

## 1. The quasi-phase-matching conditions

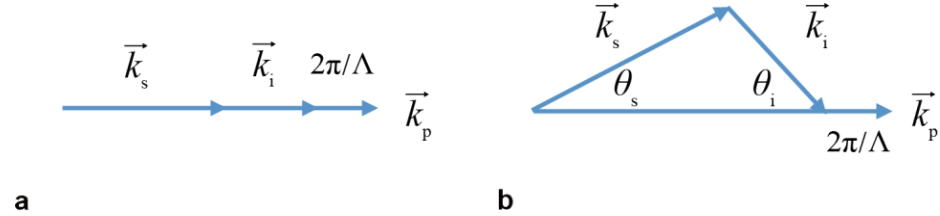

Fig. S1 | Geometric relation among the wavevectors in SPDC. **a** the collinear case; **b** the non-collinear case.

In general, the phase mismatch of quasi-phase-matching is given by [1]

$$\Delta k = k_\mathrm{p} - k_\mathrm{s} - k_\mathrm{i} - \frac{2\pi}{\Lambda} \tag{S1}$$

where $k_j = n(\omega_j)\frac{\omega_j}{c}$, $j = \mathrm{s,i,p}$; $\omega, c, n$ represent angular frequency, the speed of light, and refractive index respectively; $\Lambda$ represents the grating period of the crystal. The geometry of the wavevectors is shown in Fig. S1a. We here are interested in the non-collinear phase matching case. As shown in Fig. S1b, three wavevectors $k_\mathrm{p} - 2\pi/\Lambda$, $k_\mathrm{i}$, and $k_\mathrm{s}$ form a triangle. From the geometry, one can write the mismatch as

$$\Delta k = k_\mathrm{p} - k_\mathrm{s}\cos\theta_\mathrm{s} - k_\mathrm{i}\cos\theta_\mathrm{i} - \frac{2\pi}{\Lambda} \tag{S2}$$

When we consider the conservation of transverse momentum $k_\mathrm{s}\sin\theta_\mathrm{s} = k_\mathrm{i}\sin\theta_\mathrm{i}$ and the law of refraction $\sin\theta_\mathrm{s\_out} = n(\omega_\mathrm{s})\sin\theta_\mathrm{s}$, where $\theta_\mathrm{s}$ is the angle in the crystal, $\theta_\mathrm{s\_out}$ is the outside angle, and in a small-angle approximation the refractive index $n(\omega)$ is only dependent on wavelength and independent of angle, then the mismatch can be written as

$$\Delta k = k_p - \sqrt{k_s^2 - k_s^2 \frac{\sin^2 \theta_{s\_out}}{n^2(\omega_s)}} - \sqrt{k_i^2 - k_s^2 \frac{\sin^2 \theta_{s\_out}}{n^2(\omega_s)}} - \frac{2\pi}{\Lambda} \quad (S3)$$

## 2. Derivations from the FAS to the interference pattern

The photon number generated is proportional to a $\text{sinc}^2$ function [2]

$$C_1(\lambda_s, \theta_{s\_out}) = \alpha \text{sinc}^2\left(\frac{\Delta k \cdot L}{2}\right) \quad (S4)$$

where $L$ represents the length of crystal, $\alpha$ is a normalized factor, $\Delta k$ is given by Eq. (S3). Now, one can obtain the frequency-angular spectrum (FAS) in Fig. S2a (the same as that in Fig. 1a in the main text) by calculating and normalizing Eq. (S4). The Sellmeier equation that we use to obtain the refractive index is from ref. [3].

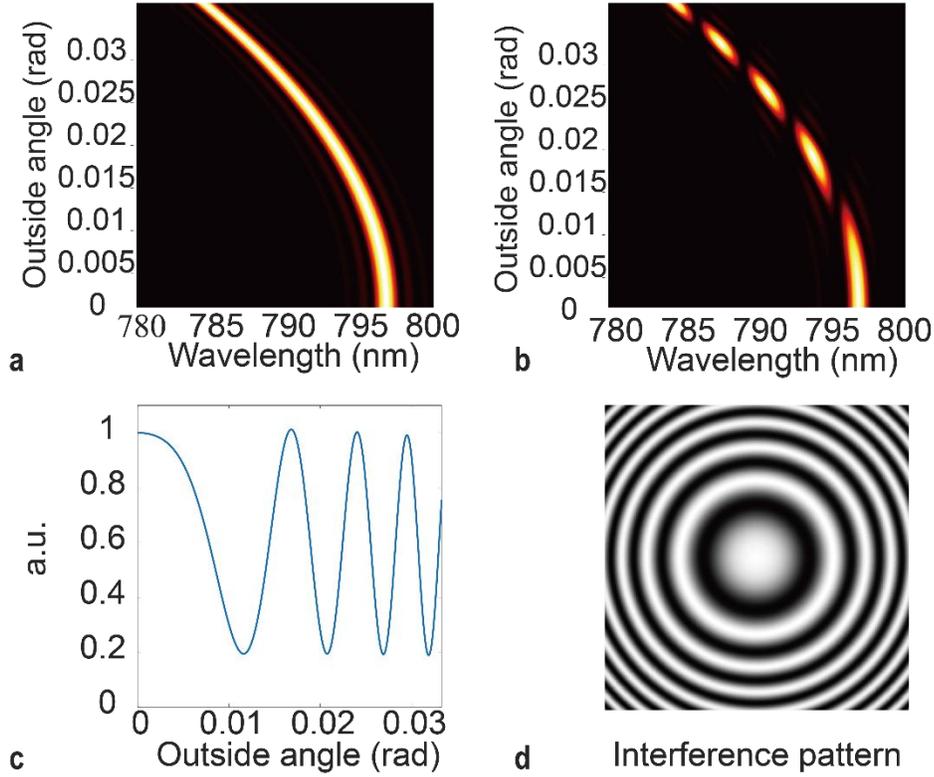

Fig. S2 | **a** /**b** The FAS before/after the interferometer. **c** The photon counts dependent on the outside angle. **d** Interference pattern with normalized intensity.

Then, we consider a Michelson interferometer shown in Fig. 2 in the main text. The divergent signal photons are collimated by a lens before the interferometer. When a monochromatic plane wave enters the interferometer, the output intensity should be multiplied by a factor $[1+\cos(4\pi d/\lambda_s)]/2$, where $d$ represents the arm difference of the Michelson interferometer and $2d$ is the optical path difference. Therefore, the photon number after the interferometer becomes

$$C_2(\lambda_s, \theta_{s\_out}) = C_1(\lambda_s, \theta_{s\_out}) \frac{1+\cos(4\pi d/\lambda_s)}{2} \tag{S5}$$

The lenses before and after the interferometer form a 4-f imaging system. On the image plane, the FAS becomes the form shown in Fig. S2b (the same as that in Fig. 2b in the main text, there the FAS is an example with $d = 100\ \mu m$). The ICCD has almost equal response for different frequency components in a small band range. In the experiment, no narrowband filter is used expect for a 750 nm long-pass filter. By integrating over the wavelength, one can obtain the photon counts dependent on the angle

$$\begin{aligned} C_3(\theta_{s\_out}) &= \int_{-\infty}^{\infty} C_2(\lambda_s, \theta_{s\_out}) d\lambda \\ &= \frac{1}{2}\left[1 + \alpha \int_{-\infty}^{\infty} \mathrm{sinc}^2\left(\frac{\Delta k \cdot L}{2}\right) \cos(4\pi d/\lambda_s) d\lambda\right] \end{aligned} \tag{S6}$$

Here, $C_3(\theta_{s\_out})$ is a single-variable function of outside angle and the integral result shows the radial counts distribution and the example numerical integrating result is shown in Fig. S2c. Considering the rotational symmetry, ring-like interference patterns can be observed on the detection plane. Assume that $C(x,y)$ represent the counts on detection plane in Cartesian coordinate, then the interference pattern, shown in Fig. S2d, can be simulated using the relation $C(x,y) = C_3\left(\sqrt{x^2+y^2}/f_3\right)$, where the approximation $\theta_{s\_out} \approx \rho/f_3$ and $\rho = \sqrt{x^2+y^2}$ are used and the integral in $C_3(\theta_{s\_out})$ should be calculated numerically. In Section 4, Eq. (S21) can be seen as an approximate integrating result.

### 3. The approximate analytical expression for the tuning curve

In this section, we deduce the approximate parabolic expression for the tuning curve in some approximation. Firstly, we approximately regard the $\mathrm{sinc}^2$ function in Eq. (2) as a delta function, this means $\Delta k = 0$. Then, we give the lengths of three sides of the triangle in Fig. S1. Finally, the angle $\theta_s$ can be written directly using the cosine theorem.

We define some auxiliary quantities for convenience

$$A = \frac{\omega_{s0} n_{s0}}{c} \tag{S7}$$

$$B = \frac{\omega_{i0} n_{i0}}{c} \tag{S8}$$

$$C = \frac{\omega_{s0}}{c}\left(\frac{dn}{d\omega_s}\right)_{\omega_{s0}} + \frac{n_{s0}}{c} = \frac{\beta_s \omega_{s0} + n_{s0}}{c} \tag{S9}$$

$$D = \frac{\omega_{i0}}{c}\left(\frac{dn}{d\omega_i}\right)_{\omega_{i0}} + \frac{n_{i0}}{c} = \frac{\beta_i \omega_{i0} + n_{i0}}{c} \tag{S10}$$

where $\Delta\omega = \omega_s - \omega_{s0} = \omega_{i0} - \omega_i$; $\beta_s = \left(\frac{dn}{d\omega_s}\right)_{\omega_{s0}}, \beta_i = \left(\frac{dn}{d\omega_i}\right)_{\omega_{i0}}$ are the first-order dispersion coefficient at the center frequency of signal and idler photons; $\omega_{s0}, \omega_{i0}, \lambda_{s0}, n_{s0}, n_{i0}$ are the center frequency and wavelength and the corresponding refractive index of signal and idler photons. Now the lengths of three sides of the triangle in Fig. S1b can be expressed as

$$k_p - \frac{2\pi}{\Lambda} = \frac{n_{s0}\omega_{s0} + n_{i0}\omega_{i0}}{c} = A + B \tag{S11}$$

$$k_{\mathrm{s}} = \frac{n(\omega_{\mathrm{s}})\omega_{\mathrm{s}}}{c} \approx \left[n_{\mathrm{s}0} + \left(\frac{dn}{d\omega_{\mathrm{s}}}\right)_{\omega_{\mathrm{s}0}} \Delta\omega\right]\frac{(\omega_{\mathrm{s}0}+\Delta\omega)}{c} \approx A + C\Delta\omega \quad \text{(S12)}$$

$$k_{\mathrm{i}} = \frac{n(\omega_{\mathrm{i}})\omega_{\mathrm{i}}}{c} \approx \left[n_{\mathrm{i}0} - \left(\frac{dn}{d\omega_{\mathrm{i}}}\right)_{\omega_{\mathrm{i}0}} \Delta\omega\right]\frac{(\omega_{\mathrm{i}0}-\Delta\omega)}{c} \approx B - D\Delta\omega \quad \text{(S13)}$$

The angle of signal photons in the crystal can be solved using the cosine theorem

$$\begin{aligned}\cos\theta_{\mathrm{s}} &= \frac{-(B-D\Delta\omega)^2 + (A+B)^2 + (A+C\Delta\omega)^2}{2(A+B)(A+C\Delta\omega)} \\ &\approx \frac{AB + A^2 + (BD+AC)\Delta\omega}{AB + A^2 + C(A+B)\Delta\omega} \\ &\approx 1 - \frac{B(C-D)}{A(A+B)}\Delta\omega\end{aligned} \quad \text{(S14)}$$

where the second-order terms including $\Delta\omega^2$ are ignored and only the linear term including $\Delta\omega$ are preserved. Considering $\cos\theta_{\mathrm{s}} \approx 1 - \frac{1}{2}\theta_{\mathrm{s}}^2$, we have

$$\theta_{\mathrm{s}}^2 = 2\left[\frac{B(C-D)}{A(A+B)}\right]\Delta\omega = 2\frac{\omega_{\mathrm{i}0}n_{\mathrm{i}0}}{\omega_{\mathrm{s}0}n_{\mathrm{s}0}}\frac{\beta_{\mathrm{s}}\omega_{\mathrm{s}0} + n_{\mathrm{s}0} - \beta_{\mathrm{i}}\omega_{\mathrm{i}0} - n_{\mathrm{i}0}}{\omega_{\mathrm{s}0}n_{\mathrm{s}0} + \omega_{\mathrm{i}0}n_{\mathrm{i}0}}\Delta\omega \quad \text{(S15)}$$

Then, considering $\theta_{\mathrm{s\_out}} \approx n_{\mathrm{s}0}\theta_{\mathrm{s}}$, we can obtain the outside angle

$$\theta_{\mathrm{s\_out}}^2 = b_1 \Delta\omega = -b_2 \Delta\lambda \quad \text{(S16)}$$

where $\Delta\lambda = \lambda_{\mathrm{s}} - \lambda_{\mathrm{s}0}$, $b_2 = \frac{2\pi c}{\lambda_{\mathrm{s}0}^2}b_1$, and

$$b_1 = \frac{2n_{\mathrm{i}0}n_{\mathrm{s}0}\omega_{\mathrm{i}0}(\beta_{\mathrm{s}}\omega_{\mathrm{s}0} + n_{\mathrm{s}0} - \beta_{\mathrm{i}}\omega_{\mathrm{i}0} - n_{\mathrm{i}0})}{\omega_{\mathrm{s}0}(\omega_{\mathrm{s}0}n_{\mathrm{s}0} + \omega_{\mathrm{i}0}n_{\mathrm{i}0})} \quad \text{(S17)}$$

### 4. The interference visibility and the coherence length

The visibility is commonly dependent on the bandwidth. Here, the function $C_1(\lambda_{\mathrm{s}}, \theta_{\mathrm{s\_out}})$ in Eq. (S4) describes the spectrum for a particular outside angle and can be approximately written in another form

$$C_1(\omega_{\mathrm{s}}, \theta_{\mathrm{s\_out}}) = \frac{1}{\delta\omega_\theta}\mathrm{sinc}^2\left[\frac{2\pi(\omega_{\mathrm{s}} - \omega_\theta)}{\delta\omega_\theta}\right] \quad \text{(S18)}$$

Where $\omega_\theta$ and $\delta\omega_\theta$ are the center frequency and the spectral full width of the FAS in Fig. S2a at the particular outside angle $\theta_{\mathrm{s\_out}}$. Then, substituting Eq. (S18) into Eq. (5) and (6), then the integrating becomes

$$\begin{aligned}C_3(\theta_{\mathrm{s\_out}}) &= \frac{1}{2}\left[1 + \int_{-\infty}^{\infty}\frac{1}{\delta\omega_\theta}\mathrm{sinc}^2\left[\frac{2\pi(\omega_{\mathrm{s}} - \omega_\theta)}{\delta\omega_\theta}\right]\cos\left(\frac{2d}{c}\omega_{\mathrm{s}}\right)d\omega_{\mathrm{s}}\right] \\ &= \frac{1}{2} + \frac{1}{2}\mathrm{tri}\left(\frac{\delta\omega_\theta}{2\pi c}d\right)\cos\left(\frac{2d\omega_\theta}{c}\right)\end{aligned} \quad \text{(S19)}$$

where the triangle function is defined as

$$\mathrm{tri}(x) = \begin{cases}1 - |x|, & -1 < x < 1 \\ 0, & \text{otherwise}\end{cases} \quad \text{(S20)}$$

Considering Eq. (S16), one can obtain $\omega_\theta = \theta_{s\_out}^2 / b_1 + \omega_{s0}$. The modified formula describing the interference becomes

$$C(\vec{\rho}; d) = \frac{1}{2} + \frac{1}{2} \operatorname{tri}\left(\frac{\delta\omega_0}{2\pi c} d\right) \cos\left(\frac{2d\omega_{s0}}{c} + \frac{2d\rho^2}{cb_1 f^2}\right)$$

(S21)

where $\rho = f\theta_{s\_out}$ is the radial coordinate on the detection plane, $f$ is the focal length of the lens before detection plane. The bandwidth $\delta\omega_\theta$ has been approximately regarded as a constant $\delta\omega_0$ that is the acceptable bandwidth in the collinear case ($\theta_{s\_out} = 0$). The visibility is given by

$$V = \frac{C_{\max} - C_{\min}}{C_{\max} + C_{\min}} \approx \operatorname{tri}\left(\frac{2d}{4\pi c / \delta\omega_0}\right) \approx \operatorname{tri}\left(\frac{2d}{2\lambda_{s0}^2 / \delta\lambda_{s0}}\right)$$

(S22)

where $\delta\lambda_{s0}$ is the full linewidth of photons in the case of $\theta_{s\_out} = 0$ and can be calculated by substituting Eq. (S1) into Eq. (S4). The triangle function describes the temporal coherence and this property can be simply reflected by a parameter, coherence length. Considering the full width of the triangle function, the coherence length is given by

$$l_c = 8\pi c / \delta\omega_0 = 4\lambda_{s0}^2 / \delta\lambda_{s0}$$

(S23)

The predicted coherence is 1.20 mm that agrees well with the experimental results shown in the main text. The linewidth is dependent on the material and length of the crystal. For a given material, the longer the crystal is the larger the coherence length is, the interference therefore requires a relatively long crystal to obtain visible stripes.

## 5. Determination of the equal path position of the Michelson interferometer

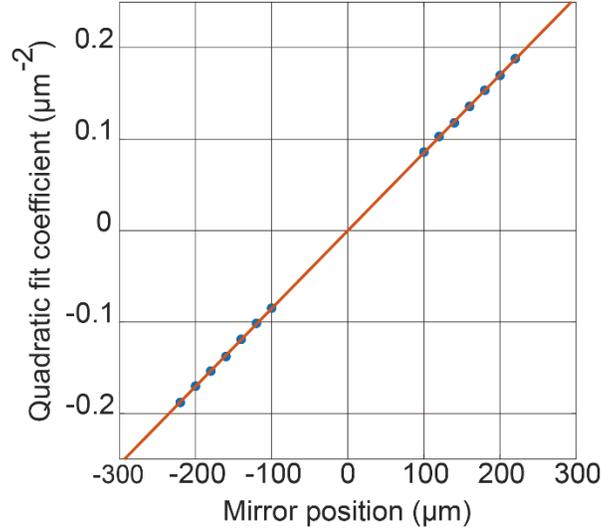

Fig. S3 | The quadratic coefficients $a$ are obtained by fitting the experimental data. The blue spots in the top right are the same as those in Fig. 3e in the main text; the blue spots in the bottom left are obtained by fitting the data recorded at $d = -220, -200, -180, -160, -140, -120, -100\,\mu\mathrm{m}$; the fitting is the same as that in Fig. 3d in the main text. The data (blue spots) are fitted by a linear function (red line).

In our experiment, the equal path position $d = 0\,\mu m$ in Fig. 3a~3c is set manually. Here, we determine the equal path position accurately by fitting the experimental data. In Fig. S3, blue spots show the quadratic coefficients at different position *d* of the mirror M2, which is fitted using a linear function $y = k(x - x_0)$. The fitting result shows the intercept $x_0$ at *x*-axis is $0.09 \pm 0.54\,\mu m$, which is the real position of equivalent path. Considering the resolution of our displacement platform is $10\,\mu m$, the error of $0.09\,\mu m$ is negligible. The uncertainty $\pm 0.54\,\mu m$ is estimated using 95% confidence bounds of the fitting.